\documentclass[letterpaper]{article}
\usepackage{digra}
\usepackage{times,graphicx,multicol}
\usepackage{fancyhdr}
\title{Modeling Epidemic Spread in Synthetic Populations ---\\ Virtual Plagues in Massively Multiplayer Online Games}

\author{Magnus Boman \\
Swedish Institute of Computer Science (SICS) and \\
The Royal Institute of Technology\\ 
Kista, Sweden\\
\texttt{mab@sics.se}
\And
\textbf{Stefan J. Johansson}\\
School of Engineering\\
Blekinge Institute of Technology\\
Ronneby, Sweden\\
\texttt{sja@bth.se}
}

\begin{document}
\thispagestyle{fancy}
\renewcommand\headrulewidth{0pt}
\cfoot{
	\framebox[\textwidth]{
\begin{minipage}{0.97\textwidth}
		{\small
			{\parbox[c]{\textwidth}{\bf\centering Situated Play, Proceedings of DiGRA 2007 Conference}}\newline\newline
			{\parbox[l]{\textwidth}{\copyright~2007 Authors and Digital Games Research Association (DiGRA). Personal and educational classrom use of this paper is allowed, commercial use requires specific permission from the author.}}
		}
	\end{minipage}
	}
}
\maketitle
\begin{abstract}
A virtual plague is a process in which a behavior-affecting property spreads 
among characters in a Massively Multiplayer Online Game (\textsc{Mmog}). 
The \textsc{Mmog} individuals constitute a synthetic population, and the game
can be seen as a form of interactive executable model for studying
disease spread, albeit of a very special kind.
To a game developer maintaining an \textsc{Mmog}, recognizing, monitoring, and ultimately 
controlling a virtual plague
is important, regardless of how it was initiated. The prospect of using 
tools, methods and theory from the field of epidemiology to do this seems
natural and appealing. We will address the feasibility of such a prospect,
first by considering some basic measures used in epidemiology, 
then by pointing out the differences between real world epidemics and virtual plagues. 
We also suggest directions for \textsc{Mmog} developer control through
epidemiological modeling.
Our aim is understanding the properties of virtual plagues, rather than
trying to eliminate them or mitigate their effects,
as would be in the case of real infectious disease.
\end{abstract}

\subsection{Author Keywords}
Virtual Plagues, Digital Plague, Infectious Disease, Epidemiology, Synthetic Population

\section{Introduction}
A virtual plague is a process in which a behavior-affecting property spreads 
among characters in a Massively Multiplayer Online Game (\textsc{Mmog}). 
The \textsc{Mmog} individuals constitute a synthetic population, and the game
can be seen as a form of interactive executable model for studying
disease spread, albeit of a very special kind.
Previous work has shown that sixth grade student players ascribe virtual 
plagues the same properties as real diseases~\cite{Neulight:2005}. 
Therefore, some players can be expected to try to use their knowledge about
disease spread when confronted with a virtual plague.

What makes the 
virtual plague special is that its initiators (be it a game developer, an
individual code developer, or a code logger) do not let
any mapping to the real world constrain them. 
In contrast to computer scientists
developing tools for studying epidemiological processes and to the physicians
and sociologists using them, the initiators seek to learn little about the 
real world. They concentrate instead on the virtual world of the game and are
happy to make the \textsc{Mmog} more interesting through the particular form of
modification (\emph{mod}) that is the virtual plague.

Depending on whether the game data is on the client side or on the server side,
virtual plague initiators (a kind of \emph{modders}) have a relatively easy or hard task, respectively, to
implement the mod and to test its effects. Some games, such as The Sims, encourage
mods (including virtual plagues) and the game has a huge
community of modders. 
In one famous case, the guinea pig mod~\cite{bbc00};
one of the first recorded virtual plagues, the culprit
turned out to be the chief game developer Mr. Will Wright himself~\cite{Markhoff:2000}.
This opened for a discussion about the responsibility from the game developer side
as to acknowledging the time and effort put in by people into their game characters,
including not only virtual plagues but also, e.g., keeping servers alive in spite
of declining numbers of players. The very first mods were made by Wizards (and players with access to the underlying text databases) of Multi-User Dungeon (MUD) in the late 1980s~\cite{Muramatsu:1998}.


The most famous example of virtual plague initiation from
the developer side is perhaps Blizzard's introduction of the Corrupted Blood debuff in
World of Warcraft. This debuff was originally conceived of as spreading from monster to
player only, although Blizzard let it transmit from player to player if the players
were standing close to each other. The idea was to let one boss monster issue the
debuff, which would then be allowed to affect all players fighting that particular monster
(and ultimately help kill off another boss monster in the same area). What
was unexpected, and seemingly a surprise to Blizzard, was that players passed the
debuff on to their pets, dismissed the pets, and then re-summoned them later, in
much more populated areas. This led to tens of thousands of deaths on at least three servers,
and caused Blizzard to try to correct the problem through various quick fixes (cf. \cite{wiki070217}).
The pet debuff trick has analogues in the real world, in viruses spreading
between humans via animal reservoirs, for instance. Epidemiologists also 
showed an interest in the human behavior connected to this virtual plague, e.g., 
people steering their characters towards large cities to pass on the disease to
as many other characters as possible before they died~\cite{syd05}.

Another example is the \emph{gray plague} in the game \emph{Kingdom of Loathing} 
which was initially spread from contaminated \emph{comfy blankets},
no doubt a pointer to the infamous (and some claim deliberate) 
spread of smallpox to the native Americans in the siege of Fort Pitt in 1763. 
In the game, the plague spread via the chat and the messaging between the players, 
regardless of player positions on the map. This has no correspondance to real world
spread of disease, but rather to rumor diffusion. 
Since the chat and the message passing are the main tools for networking, 
it was a successful channel for spread. After a few weeks, 
a serum was given to the players in a special quest, which wiped out the plague~\cite{kolwiki:2007}.

To a game developer maintaining an \textsc{Mmog}, recognizing, monitoring, and ultimately 
controlling a virtual plague
is important, regardless of how it was initiated. The prospect of using 
tools, methods and theory from the field of epidemiology to do this seems
natural and appealing. We will address the feasibility of such a prospect,
first by considering some basic measures used in epidemiology, 
then by pointing out the differences between real world epidemics and virtual plagues. 
We will finally suggest directions for \textsc{Mmog} developer control through
epidemiological modeling. 
Our aim is understanding the properties of virtual plagues, rather than
trying to eliminate them, protect individuals, or mitigate their effects,
as in the case of real infectious disease. This is in part due to our stance that the effects of a virtual plague are not necessarily unwanted. We will not discuss computer viruses, since these are already well understood, but more importantly they are an example of a virtual phenomenon crossing the bridge to the real world. We therefore stick to the mapping between on the one hand a purely synthetic population subjected to a virtual plague and,on the other hand, a real population subjected to epidemiological studies pertaining only to real disease.

\section{Basic Epidemiological Measures}
The classical \textsc{Sir} model~\cite{anma91} divides the affected population into three stages:
Susceptible, Infected, and Recovered/Removed. It is a top-down macro model based on homogeneous
mixing: you essentially throw a population, with relevant distributions, as well as a
pathogen and some of its properties, into a bag. Then you shake the bag, and an affected population results.
No explicit modeling of space is done, and no outliers (such as super-spreaders) are considered;
the model is built on averages, and may be described in its entirety by a set of differential
equations.

Micro-models, by contrast, let a set of individuals populate a synthetic landscape,
and then execute a simulation run, in which an infectious disease spread pattern emerges.
Due to stochastic elements, this pattern varies with each run, and so simulations are
often repeated over long series of runs. In particular, many runs might be required before
anything interesting happens, since realistic parameters will imply that an epidemic does not actually
occur very often. Each individual has a state, i.e. the individuals are agents part of a multi-agent
system (cf. \cite{boho04}). In particular, the state carries information about the progress of the disease in
those agents that are affected by the pathogen. The smallpox model of 
Halloran \emph{et al.}~\cite{haloniya02} may serve as an example here:
\begin{enumerate}
	\item Incubating, noninfectious, vaccine-sensitive (first 3 days after infection)
	\item Incubating, noninfectious, vaccine-insensitive (remaining 7-11 days)
	\item Prodromal, highly infectious (3-5 days)
	\item Symptomatic, infectious (10 \% of Prodromal stage), withdrawal (over first 3 days out of 14-17)
	\item Recovered/removed
\end{enumerate}

These stages of disease were used for each individual in a discrete time
microsimulation based on stochastic generation of $2000$ individuals, based on US census data.
This is essentially a Markov-chain model, typical for simulating epidemic spread bottom-up.
Time is crucial, and the most ambitious models of this kind also have an explicit representation
of space (see, e.g., \cite{eub04,bro05}). This means that an entirely new level of adequacy with respect to reality can be
reached in the modeling, as compared to the classical macro models. 

The \emph{basic reproductive rate} $R_{0}$ is often used, in macro- as well as micromodels.
In spite of its popularity among modelers, it is a problematic semi-formal aggregate measure.
Intuitively, it is the average number of individuals
directly infected by an infectious case during his or her entire infectious period,
when he or she enters a totally susceptible population (see, e.g., \cite{gi02}).

There are two ways of coding the $R_{0}$ of the disease: micro- or macrocoding.
In the macrocoding, the rate is set \emph{ex ante}, based on empirical evidence
from earlier outbreaks of the disease under study.
The $R_{0}$ value assignment is 
complicated by the fact that one may wish to separate first-generation cases from
later-generation cases, e.g., by taking the weighted average. The rate
is also contingent on the rate of vaccination, further complicating macrocoding.
Assumptions of a high rate of vaccination reduce the $R_{0}$ value, for instance,
but making good guesses about the actual rate of vaccination is difficult,
for many diseases.

In the microcoding, $R_{0}$ is potentially different in each simulation run
and any description of the emergence of the value will be made \emph{ex post}. The
purpose of \emph{ex post} descriptions is to verify that the rate does not reach absurd size,
or that it does not vary too much between various geographical regions, modulo population
density~\cite{botofl99}.

\section{Epidemiology for Virtual Plagues}
Keeping a virtual plague in check is very different from keeping real infectious disease
in check. In particular, it is a lot easier, and most importantly its effects are 
negligable compared to those of real disease. But there are other intriguing aspects
of virtual plagues. They might provide for more interesting game play, but only
if they are neither too devastating, nor too easy to fend off. In order to keep
virtual plagues active, a macro modeler would say that the $R_{0}$ value should be
very close to $1$. A micro modeler would rather say that the virtual plague should
spread at just the right pace, in interesting and unpredicted patterns.

A large number of reasons make micromodeling virtual plagues a better alternative
than macromodeling them:
\begin{itemize}
\item
First, there is no empirical evidence to build on.
Earlier outbreaks of the same virtual plague, if they have occurred at all,
will either present a solved problem (i.e., the bug has been fixed, or the players
know how to fix the problem in-game, e.g., by completing a new quest) or an entirely new problem (i.e., no player
will know what to do). 
\item
Second, the (either geographical or virtual) location of players in-game is important,
just as in real life epidemics. Even if players can teleport between certain places
in certain games, the distance usually yields a constraint for disease transmission
which may be modeled in micro, but not in the classical macro models. 
\item
Third, players
are heterogeneous. The capability to heal, for instance, varies greatly between 
players, even within the same vocation. The analogue to real life resistance to a 
virus, possibly due to a vaccination earlier in life is tempting but unreasonable,
since the variance among individuals is much larger than in human populations,
for any disease. At any rate, the homogenous mixing utilized in the macro approach will
not work well to mirror \textsc{Mmog} spread.
\item
Fourth, the aims and intentions of the players can not be captured in macro modeling. 
Since the cost of being infected by a disease in most games is negligible, 
players happily enough tend to be much more risk averse in real life than in games. 
In real life, there are two diffusion patterns that need to be modeled in the case
of an epidemic: the diffusion of the disease itself, and the diffusion of information
about the disease. These two patterns affect each other. 
For instance, in real life, people that are aware of the fact that the hospital is a high-risk place to be during an epidemic, will not go there if they catch a high fever, because they might catch a much more dangerous disease if they go to hospital with the intent
of being examined by a doctor. 
In a game, curiosity is likely to instead drive
people towards the epicenter of a virtual plague, because that is where the game
action currently is. This is precisely what happened in the case of Corrupted Blood,
when people gathered in the city of Ironforge, for example.
Generally speaking, estimating the effect of high-risk behaviors, such as through logging the actions by super-spreaders,
or the intentional spread by so-called zombies, is much easier in micro than in macro.
\item
Fifth, the ability for virtual plagues to spread through different channels is not easily modeled 
in macro. However, it may be necessary to be able to model all aspects of the 
virtual plague in order to understand how to halt its spread. 
The more channels that the plague may spread through, the harder it is to control it by,
for example,
limiting the use of a chat, or preventing players from entering a contaminated area in the game.
\end{itemize}

\section{Controlling the Impact of Virtual Plagues}
The impact of a virtual plague on any \textsc{Mmog} is very difficult to control. However, there are a number of possibilities that could be used by developers or that people with server code access could exploit. In the best of all worlds, this leads to a \textsc{Mmog} benefiting from interesting emergent behaviour-changing patterns, from which there is something to be learned for all players affected. If this is to hold true, virtual plague initiators should be aware of a number of possible pitfalls and misconceptions. We try to provide the start of such a catalog here. The items in our catalog can hopefully
serve as guidelines for controlling social and even economic effects of badly engineered virtual plagues.

\subsection{The Initial Phase}
The very first days of an outbreak of a potential epidemic are vital to policy makers and epidemiologists studying real disease. This is because the policy measures set in can actually determine whether there will be an epidemic at all. The models for studying initial spread are also different from the models used later, when the disease is a pandemic or is endemic. In the case of virtual plagues, if there is no warning and no information whatsoever about what is going on, players will spread false rumors about the disease. 
These rumors will propagate quickly, which is not necessarily bad for game play, but will lead to players responding differently to the virtual plague. 
In the \textsc{Mmog} Tibia, one or more warnings preceed each monster raid, described by game developer Cipsoft as \emph{automated events}~\cite{Tibia:2007}. 
These warnings are different for each of the 60 different raids, sending players to approximately the right place and with approximately the right expectations. Something similar is recommended for virtual plagues. Will players heed the warnings? Not necessarily, but at least the first players encountering the virtual plague will be unfearing high levels, if the information issued is correct.

\subsection{The Epidemic Phase}
Once the virtual plague is past its first days since introduction, some places should be heavily affected and others unaffected. One way of controlling player behavior is to 
restrict access to heavily hit areas. Players will most likely try and go there in spite of restrictions, creating new areas of intensely populated areas, just outside the restricted area. These are good candidate areas for spreading more information about the virtual plague, as most players will be there with the purpose of finding out more about the disease.

During the period of maximal intensity of the virtual plague, i.e. when the most players are being infected, the adequacy of parameter settings for the individual virtual plague stages will be tested to the full. If for example the incubation period is too long, or if the transmission probabilities are too high, this will show, e.g. through too many players dying or being rendered immobile by the virtual plague (somewhat counter-intuitively possibly leading to it to die off, since there will not be enough new victims for it to spread!). Experimenting with individual costs, e.g., letting individuals be affected differently, either contingent on their state when they contracted the virtual plague, or on stochastic variables, is also important, as it further underlines heterogeneity. 
For instance, healthy individuals could behave as if they had immunity, as could individuals that have already suffered from the virtual plague at least once. Decisions have to be made about whether or not immune individuals spread the virtual plague, and for how long. In the case of real diseases, the most feared ones are generally speaking those in which victims feel well enough to move around while they are contagious, and get notably sick only in later stages of the disease.
Non-immune individuals that have contracted the virtual plague could show this in appearance or behavior, e.g., by moving slower or less, as if "drunk" or as if affected by a spell. This was not carried out to enough extent in the case of Corrupted Blood, which was also much too contagious and arguably much too harmful.

\subsection{The Recovery Phase}
The solution to players combatting a virtual plague could be a virtual cure, the finding of which makes for an excellent quest. Much like the scientific community relatively successfully put aside competition in order to swiftly cooperate to control the real life \textsc{Sars} epidemic, one could wish for players to set aside differences like guild wars and revenge schemes in order to fight this new threat. 
In this way, a virtual plague could be a bringer of peace to a \textsc{Mmog}. This, however, requires the quest to be specified somehow, at least in the sense that sufficient clues could be obtained before too much suffering from the virtual plague was endured. Introducing a new boss monster would be one example. 

\subsection{The Case of the Gray Plague}
As an example of the three phases described above, let us see what happened in the case of the Gray Plague in the Kingdom of Loathing.
\subsubsection{The Initial phase}
The Gray Plague showed its first symptoms in the chat three days after the first players got infected. Someone noticed, and was able to confirm through image analysis in \textsc{Gimp}, that the usually black text in the chat had turned dark gray. A \emph{-cough-} was also added now and then when the infected player chatted. 

\subsubsection{The Epidemic Phase}
As the text got lighter and lighter gray, and players also began to shudder, the players looked for cures, 
for instance a pair of rose-colored glasses that made the text look black again to the player. 
However, the glasses merely took away the symptoms, and the player still infected chat mates. 
Another attempt was made to use \emph{Doc Galactic's Medical Cure}. It temporarily cured the player, but the player was still susceptible to the plague if it entered the chat again or got in-game messages from infected players. Since there was no coordinated action, the cure was not effective.

\subsubsection{The Recovery Phase}
\begin{figure}
\begin{center}
\includegraphics[width=0.9\columnwidth]{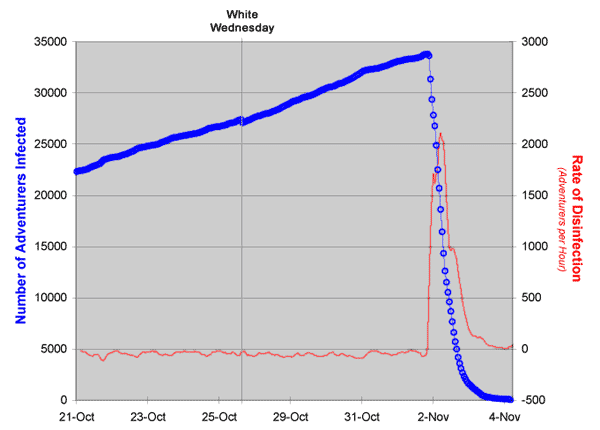}
\end{center}
\caption{The spread of the Gray Plague in the Kingdom of Loathing.}
\label{grayplague}
\end{figure}

Finally, the council set up a quest in which the players could get the items needed for making a serum blowgun that made them immune to the gray plague. The plague was more or less wiped out in three days, see Figure~\ref{grayplague}~\cite{kolwiki:2007}.

\subsection{Evolutionary Development}
A virtual plague could mutate, like viruses do in the real world. In this way, it could tacitly remain in the \textsc{Mmog} for months or even years, unbeknownst to many players, and then reveal itself, either as a deadly plague, or as something contagious but quite harmless (e.g., producing scars on the avatars, or requiring medicine to regain stats).

\subsection{Learning from Virtual Plagues}
Browsing blogs from the weeks after the Corrupted Blood event, one finds multiple comments worrying about Corrupted Blood being intentionally spread by high-level players, with the malicious intent of studying effects of the virtual plague: The ulterior motive of those players was by the commentators supposed to be the unleashing of a disease in real life, which could be made more devastating thanks to what they learned from the virtual plague. Not only was this not the case with Corrupted Blood, since it was initiated, albeit not entirely controlled at all times, by Blizzard. It is also an unrealistic scenario because of the complexity of the game. If fact, any \textsc{Mmog} is likely to be a much too complex system to allow for studies with implications for real disease. There are several simulation environments that lend themselves to such malicious intent, and even if the community awareness of terrorism and similar serious threats among computer scientists creating these environments is high, terrorists would much more likely try and get their hands on such simulation platforms than manipulating \textsc{Mmog}s. 
There are, however, a number of properties that can (and in the case of virtual plagues maybe even should)
be shared between real-life epidemics and virtual plagues:

\begin{itemize}
\item
Both domains need to model individuals situated in an environment 
that allows for the transfer of the disease between them, 
subject to constraints coded by various transmission probabilities called \textsc{Beta}-values.
Occasionally, the environment itself (in the case of virtual plagues through pets or \textsc{Npc}s, for instance)
may host the disease, on its way between two individuals.
\item
The disease does (in the vast majority of cases) decrease the capabilities of the individuals in some way.
A virtual plague may cause the death of a character, which may cause the player much distress,
depending on the in game penalty for death (banishment, loss of eq, mana, monetary means, exp, etc.).
\item
Measures may be taken by the individuals to decrease the risk of being infected.
\end{itemize}
However, the possibility to handcraft virtual diseases also opens up for new virtual plague designs, 
violating the classical physical laws that real diseases have to obey. 
Virtual plagues spread through chat and message channels is one such novel approach that, 
even though physical contact in the virtual world is avoided, still can be justifiable in the game.
Here sociological research into the spread of rumors is more relevant than epidemiology,
if scientific explanations of gaming behavior are sought (cf., e.g., \cite{no01}).

\section{Conclusions}
We have presented guidelines for controlling, and to some extent for designing, virtual plagues in \textsc{Mmog}s.
Microsimulation is currently the only way to adequately model virtual plagues, and we have shown why
the suggested use of traditional epidemiological models from the \textsc{Sir} family is inappropriate.
The study of virtual plagues is still in its infancy, but if game data is made available
this would enable more formal studies of their properties, which vary between different \textsc{Mmog}s.
Ambitious micromodels of real infectious disease could be of pivotal importance to such studies.

\section{Acknowledgements}
The authors would like to thank the Swedish Institute of Computer Science and Blekinge Institute of Technology for their support of our work.

\bibliographystyle{plain}
\bibliography{DiGRA07}
\end{document}